\documentclass[%
floats,aps,showpacs,twocolumn,superscriptaddress,pre,reprint,floatfix
]{revtex4-1}

\usepackage{graphicx}
\usepackage{amsmath}   
\usepackage{amsfonts}
\usepackage{nicefrac}   
\usepackage[english]{babel}

\providecommand*{\p}{\ensuremath{p}}
\providecommand*{\q}{\ensuremath{q}}
\providecommand*{\Ndof}{\ensuremath{f}}

\providecommand*{\xic}{\ensuremath{\xi_{12}}}

\providecommand*{\pointPSname}{\ensuremath{u}}
\providecommand*{\pointPS}{\ensuremath{\vec{\pointPSname}}}

\providecommand*{\sectionnormal}{\ensuremath{n}}
\providecommand*{\vecsectionnormal}{\ensuremath{\vec{\sectionnormal}}}
\providecommand*{\sectiondist}{\ensuremath{D}}

\providecommand*{\periodicpoint}{\ensuremath{{\pointPS_{\mathrm{p}}}}}
\providecommand*{\fixedpoint}{\ensuremath{{\pointPS_{\mathrm{fp}}}}}

\providecommand*{\oneD}{\textsc{1d}}
\providecommand*{\twoD}{\textsc{2d}}
\providecommand*{\threeD}{\textsc{3d}}
\providecommand*{\fourD}{\textsc{4d}}
\providecommand*{\fiveD}{\textsc{5d}}
\providecommand*{\sixD}{\textsc{6d}}
\providecommand*{\fD}[1][\Ndof]{{#1}\textsc{d}}

\providecommand*{\FLI}{\textsc{fli}}

\providecommand*{\ATLAS}{\textsc{atlas}}
\providecommand*{\mayavi}{M\textsc{ayavi}}

\providecommand*{\sectioneps}{\ensuremath{\varepsilon}}

\providecommand*{\symbolsection}{\ensuremath{\Gamma}}
\providecommand*{\Geps}[1][\sectioneps]{\ensuremath{\symbolsection_{#1}}}

\providecommand*{\colorresonance}{orange}

\providecommand*{\colorresonanceb}{brown}

\providecommand*{\colorregularbeyondabent}{pink}

\providecommand*{\colorregularcenter}{red}

\providecommand*{\colorregularbeyonda}{violet}

\providecommand*{\colorregularbeyondb}{cyan}

\providecommand*{\colorresonancethreetower}{green}

\providecommand*{\psslc}{\threeD{} phase-space slice}
\providecommand*{\psslcs}{\psslc s}
\providecommand*{\Psslc}{\psslc}
\providecommand*{\Psslcs}{\Psslc s}

\providecommand*{\subfiga}{(a)}
\providecommand*{\subfigb}{(b)}
\providecommand*{\subfigc}{(c)}
\providecommand*{\subfigd}{(d)}
\providecommand*{\subfige}{(e)}
\providecommand*{\subfigf}{(f)}

\ifx\theResA\undefined
\newcounter{ResA}
\fi
\ifx\theResE\undefined
\newcounter{ResE}
\fi

\newlength{\subfigurewidth}
\usepackage{array}
\newcolumntype{P}[1]{>{\centering\let\newline\\\arraybackslash\hspace{0pt}}p{#1}}
\newcolumntype{M}[1]{>{\centering\let\newline\\\arraybackslash\hspace{0pt}}m{#1}}

\newcolumntype{L}[1]{>{\raggedleft\let\newline\\\arraybackslash\hspace{0pt}}b{#1}}

\usepackage{amsfonts}

\providecommand*{\Z}{\mathbb{Z}}

\providecommand*{\volumeunitcellfD}{\ensuremath{V^{\text{unit cell}}_\text{\fD}}}

\renewcommand{\Re}{\ensuremath{\mathrm{Re\ }}}
\renewcommand{\Im}{\ensuremath{\mathrm{Im\ }}}

\providecommand*{\DPeigenvalue}{\ensuremath{\lambda}}
\providecommand*{\DPevec}{\ensuremath{\chi}}

\providecommand*{\eigenphase}{\ensuremath{\varphi}}
\providecommand*{\bra}[1]{\ensuremath{\langle {#1} |}}
\providecommand*{\ket}[1]{\ensuremath{| {#1} \rangle}}
\providecommand*{\braket}[2]{\ensuremath{\langle {#1} | {#2} \rangle}}

\providecommand*{\psiqnums}[2][\psi]{\ensuremath{{#1}_{#2}}}

\providecommand*{\Uopsymb}{\ensuremath{U}}
\providecommand*{\Uop}{\ensuremath{\hat{\Uopsymb}}}
\providecommand*{\Husimi}{\ensuremath{\mathsf{H}}}
\providecommand*{\coherent}{\ensuremath{\text{coh}}}

\providecommand*{\varqn}[1][n]{\vec{\q}_{\vec{#1}}}
\providecommand*{\varpj}[1][j]{\vec{\p}_{\vec{#1}}}

\providecommand*{\skewsecnamex}{\ensuremath{\alpha}}
\providecommand*{\skewsecnamey}{\ensuremath{\beta}}
\providecommand*{\skewsecnamez}{\ensuremath{\gamma}}
\providecommand*{\skewsecnamew}{\ensuremath{\delta}}

\providecommand*{\vecealpha}{\ensuremath{\vec{e}_\skewsecnamex}}
\providecommand*{\vecebeta}{\ensuremath{\vec{e}_\skewsecnamey}}
\providecommand*{\vecegamma}{\ensuremath{\vec{e}_\skewsecnamez}}
\providecommand*{\vecedelta}{\ensuremath{\vec{e}_\skewsecnamew}}

\providecommand*{\dualvecealpha}{\ensuremath{\vec{\tilde e}_\skewsecnamex}}
\providecommand*{\dualvecebeta}{\ensuremath{\vec{\tilde e}_\skewsecnamey}}
\providecommand*{\dualvecegamma}{\ensuremath{\vec{\tilde e}_\skewsecnamez}}
\providecommand*{\dualvecedelta}{\ensuremath{\vec{\tilde e}_\skewsecnamew}}

\newcounter{resstructcolumncounter}

\providecommand*{\heff}[1][]{\ensuremath{h^{#1}_{\text{eff}}}}
\providecommand*{\hbareff}{\ensuremath{\hbar_{\text{eff}}}}
\providecommand*{\ue}{\text{e}}
\providecommand*{\ui}{\text{i}}
\usepackage{color}

\providecommand*{\QMdimperdof}{\ensuremath{N}}

\providecommand*{\QMdim}{\ensuremath{\QMdimperdof^2}}

\providecommand*{\ioverhbar}{\frac{\ui}{\hbareff}}

\providecommand*{\Lanczosinitial}{\ensuremath{\chi}}

\newcommand{\MOVIEREF}{%
  For a rotating view see the supplemental material \cite{supplementalvideos}.}

\newcommand{\MOVIEREFALT}{%
  For a rotating view of each picture and a video showing the
  variation of $\p_2^*$ see the supplemental
  material \cite{supplementalvideos}.}

\usepackage{hyperref}

\begin{document}
\title{Visualization and comparison of classical structures and quantum states of 4D maps}

\author{Martin Richter}
\author{Steffen Lange}
\author{Arnd B\"acker}
\author{Roland Ketzmerick}

\affiliation{Technische Universit\"at Dresden,
  Institut f\"ur Theoretische Physik
  and Center for Dynamics,
  01062 Dresden, Germany}
\affiliation{Max-Planck-Institut f\"ur Physik komplexer Systeme,
  N\"othnitzer Stra{\ss}e 38,
  01187 Dresden, Germany}

\date{\today}

\begin{abstract}
  For generic \fourD{} symplectic maps we propose the use of \psslcs{}
  which allow for the global visualization of the geometrical
  organization and coexistence of regular and chaotic motion. As an
  example we consider two coupled standard maps. The advantages of the
  \psslcs{} are presented in comparison to standard methods like
  \threeD{} projections of orbits, the frequency analysis, and a chaos
  indicator. Quantum mechanically, the \psslcs{} allow for the first
  comparison of Husimi functions of eigenstates of \fourD{} maps with
  classical phase space structures. This confirms the semi-classical
  eigenfunction hypothesis for \fourD{} maps.
\end{abstract}


\maketitle


\section{\label{sec:intro}Introduction}

%
Understanding higher-dimensional, dynamical systems, even with just a
few particles, is a challenging task. Such systems are relevant in
many areas of physics and chemistry~\cite{LicLie92}
ranging from the dynamics of the
solar system~\cite{UdrPfe1988, Las1990, Cin2002},
dynamics of particle accelerators~\cite{DumLas1993}
to atoms and molecules~\cite{RicWin1990, SchBuc1999, Kes2007,WaaSchWig2008}.
Particular topics of interest concern the quantum signatures
of Arnold diffusion \cite{Shu1976, DemIzrMal2002a, MalChi2010}
and quantum-classical correspondence in higher-dimensional mixed
systems~\cite{AdaTodIke1988, GadReeKriSch2013}.

A standard example are autonomous Hamiltonian systems with three
degrees of freedom which have a \sixD{} phase space, which can be
reduced to a \fiveD{} manifold due to energy-conservation. Introducing
a Poincar\'e section leads to a \fourD{} symplectic map. This type of
map also arises from time-periodically driven, Hamiltonian systems
with two degrees of freedom, where a stroboscopic Poincar\'e section
leads to a symplectic map acting on the \fourD{} phase space. Such
\fourD{} maps are prototypical for the behavior of higher-dimensional
systems, as they have the smallest possible dimension showing the
essential difference to \twoD{} symplectic maps: In \twoD{} maps tori
are one-dimensional and thus lead to absolute barriers of motion. In
contrast in \fourD{} maps regular tori are two-dimensional manifolds
in the \fourD{} phase space which cannot divide the phase space into
dynamically distinct regions. One of the most fundamental consequences
of this topological structure of \fourD{} (and higher-dimensional)
maps is that generically chaotic orbits can get arbitrarily close to
any point in phase space, even if regular tori are present. One
possible mechanism was constructed by Arnold~\cite{Arn1964} leading to
the so-called Arnold
diffusion~\cite{Chi1979,KanBag1985,LicLie92,Cin2002}.
Another striking phenomenon is the occurrence of power-law trapping of
chaotic orbits in higher-dimensional systems with a mixed phase
space~\cite{DinBouOtt1990, AltKan2007, She2010, Lan2012}, for which
the mechanism is still an open question.
While most of the analytical and numerical results about
higher-dimensional systems have been obtained for the near-integrable
case, many practical applications are concerned with generic systems,
which cannot be described by perturbative methods, see, e.g.,
Refs.~\cite{GekMaiBarUze2006,PasChaUze2008,LuqVil2011} and references
therein.

For the lowest-dimensional Hamiltonian systems with regular and
chaotic dynamics, like \twoD{} billiards or time-periodically driven
\oneD{} systems, the dynamics can be reduced to \twoD{} symplectic
maps. Their phase space can be easily visualized, providing
substantial insight and intuition of the dynamics in phase space, like chaotic
motion, regular regions formed by \oneD{} regular tori around stable
periodic orbits, stable and unstable manifolds of unstable periodic
orbits, nonlinear resonances, and hierarchical regions due to partial
barriers at the border of the regular island. Also the time-evolution
of trajectories can be visualized straightforwardly.

Such a direct visualization of the classical dynamics in phase space
is also very useful when trying to understand the properties of the
corresponding quantum mechanical system. According to the
semiclassical eigenfunction hypothesis~\cite{Per73, Ber1977b, Vor79}
one expects that eigenstates semiclassically concentrate on those
regions in phase space which a generic orbit explores in the long-time
limit. For ergodic systems this is proven by the quantum ergodicity
theorem stating that almost all eigenfunctions become equidistributed
in the semiclassical limit~\cite{Shn74, Zel87, CdV85}. For systems
with a mixed phase space one thus expects that (almost all)
eigenstates either concentrate in the chaotic region or within the
regular regions on the invariant tori. Away from the semi-classical
limit this can be violated due to dynamical
tunneling~\cite{HufKetOttSch2002,BaeKetMon2005,BaeKetMon2007,IshTanShu2010}
and partial transport barriers~\cite{MicBaeKetStoTom2012}. The
semi-classical eigenfunction hypothesis has been confirmed in several
studies for \twoD{} billiard systems and maps, see
e.g.~\cite{FriDor97, VebRobLiu1999, Bae2003, BaeFueSch2004,
  MarKeeZel2005}\nocite{DegGra2003} and references therein. In order
to study this question for \fourD{} maps, it is necessary to visualize
the organization of phase space and in particular to display sets of
individual tori.

\begin{figure}
  \includegraphics{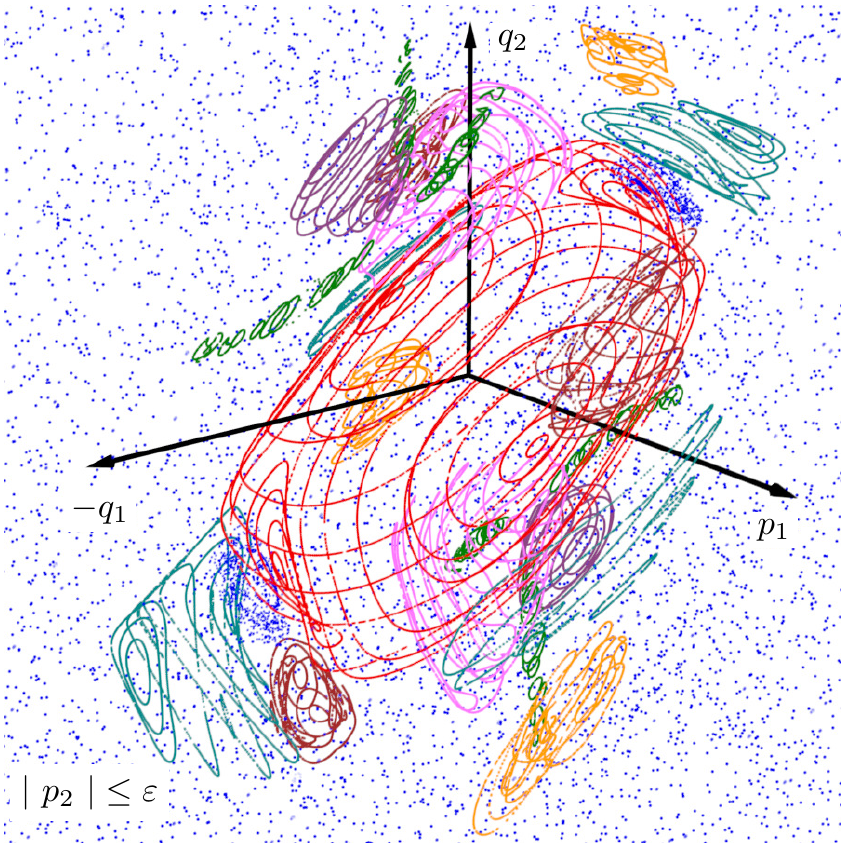}
  \caption{\label{fig:psslice-csm-strong-coupling} %
    \Psslc{} $|\p_2 | \le \sectioneps$ of the \fourD{} map given by
    two coupled standard maps, Eq.~\eqref{eqn:CoupledStdMaps}, at
    strong coupling $\xic=1$. Regular tori are given by \oneD{} lines
    (\colorregularcenter, \colorregularbeyondb, \colorregularbeyonda,
    \colorregularbeyondabent, \colorresonancethreetower,
    \colorresonance, and \colorresonanceb). They are surrounded by
    chaotic orbits (blue dots). The central regular region
    (\colorregularcenter) around the elliptic-elliptic fixed point is
    predominantly filled with regular tori (not shown for visual
    reasons). %
    \MOVIEREF }
\end{figure}

%
%
For higher--dimensional systems a direct visualization of phase space
is not possible. Starting with the pioneering work of
Froeschl\'e~\cite{Fro1970a, Fro1972} several methods have been introduced
to obtain a reduction to understand the dynamics. For example
two-dimensional plots of multi-sections~\cite{Fro1972, FroLeg2000}
or projections to two~\cite{Fro1970a, UdrPfe1988, SkoConPol1997}
or three~\cite{MarMag1981, Mag1982, VraBouKol1996} dimensions,
also including color to indicate the projected
coordinate~\cite{PatZac1994, KatPat2011},
frequency analysis~\cite{MarDavEzr1987, Las1993, ChaWigUze2003, SetKes2012},
and action-space plots~\cite{BazBonTur1998}
\label{text:numerical-methods-from-literature}. %
Further tools to investigate higher-dimensional phase spaces are chaos
indicators to distinguish regular from chaotic motion, like
finite-time Lyapunov exponents~\cite{BenGalGioStr1980a,Sko2010,ManBeiRos2012},
Fast Lyapunov Indicator~\cite{FroLegGon1997, FroGuzLeg2000, FroLeg2000},
and many more, see e.g.~\cite{Sko2001,MafDarCinGio2011, Zot2012}
and references therein.

The aim of this paper is to obtain a direct visual approach for an
understanding of \fourD{} symplectic maps similar to the insights
already available for \twoD{} maps. For this we propose the use of
\psslcs{}, see Fig.~\ref{fig:psslice-csm-strong-coupling} for an
illustration of regular \twoD{} tori and one chaotic orbit. We make
use of the full dimensionality of the \threeD{} slice by displaying
and rotating it using standard \threeD{} graphics. Since it is very
instructive to see this rotation, for the convenience of the reader
for all \psslcs{} in this paper videos with rotating camera position
are given in the supplemental material~\cite{supplementalvideos}. The
reduction from \fourD{} to \threeD{} is obtained by selecting those
points of an orbit which lie in the \psslc{} and plotting the
remaining $3$ coordinates. Thereby, \psslcs{} allow for displaying
several different orbits at the same time and provide a global
visualization of their arrangement in phase space. By this we are able
to visualize quantum eigenstates of a generic system far away from
integrability and compare them to the classical phase-space
structures. This confirms the semi-classical eigenfunction
hypothesis for regular and chaotic states in the case of \fourD{}
maps. Moreover, a combination of the \psslcs{} and the frequency
analysis enables to identify regular subregions and their separating
resonance gaps. This also explains the nature of the so-called tube
tori~\cite{VraBouKol1996,KatPat2011}.

This paper is organized as follows: In
Sec.~\ref{sec:classical-phase-space} we introduce the \psslcs{} and
the generic \fourD{} symplectic map. The observations are compared
with complementary methods, namely \threeD{} projections, frequency
analysis, and the Fast Lyapunov Indicator. In section
Sec.~\ref{sec:quantum-mechanics} we consider the quantum mechanics of
\fourD{} maps and the visualization of eigenstates by means of the
Husimi function on the \psslcs{}. Finally, Sec.~\ref{sec:conclusion}
gives a summary and an outlook. Additionally, we discuss different
slice conditions and combine \psslcs{} with a normal form
transformation in the Appendix.

\section{Regular phase space structures}
\label{sec:classical-phase-space}

\subsection{\label{sec:sections-through-ps} 3D phase-space slices}

While the concept of \psslcs{} applies to arbitrary maps, we restrict
our presentation to symplectic maps. For a symplectic map acting on a
$2\Ndof$-dimensional phase space an orbit started at an initial point
leads to a sequence of points $(p_1, \dots, p_\Ndof, \q_1, \dots,
\q_\Ndof)$. For $\Ndof \ge 2$ the dynamics cannot be displayed
completely in a \twoD{} or \threeD{} figure. To obtain a visualization
of the dynamics of higher-dimensional systems we use of \psslcs{}
$\Geps$, which are defined by thickening a \threeD{} hyperplane
\symbolsection{} in the $2\Ndof$-dimensional phase space. To specify
\symbolsection{} the simplest choice is to fix $2\Ndof - 3$
coordinates from $p_1, \dots, p_\Ndof, \q_1, \dots, \q_\Ndof$. For
\fourD{} maps ($f=2$) one thus has to fix one coordinate, e.g.\ $\p_2
= \p_2^*$, to define the slice by
\begin{align}
  \label{eq:slice-condition-in-coordinate}
  \Geps &= \left\{ (\p_1, \p_2, \q_1, \q_2) \; \left| \rule{0pt}{2.4ex} \;
         |\p_2 - \p_2^*| \le \sectioneps \right. \right\}.
\end{align}
Whenever a point of an orbit lies within $\Geps$, the remaining
coordinates $(\p_1, \q_1, \q_2)$ are displayed in a \threeD{} plot.
This is different to the method by Froeschl\'e~\cite{Fro1970a,Fro1972}
which first projects onto the \threeD{} phase space and then uses
\twoD{} slices to visualize this projection. Slice conditions like
Eq.~(\ref{eq:slice-condition-in-coordinate}) have been considered, e.g.,
in Ref.~\cite{FroLeg2000}, but without \threeD{} visualization.

By the parameter $\sectioneps$ the resolution of the resulting plot is
controlled. Decreasing $\sectioneps$ gives a tighter slice condition
but also requires to numerically compute longer trajectories as the
slice condition is fulfilled less often. This also shows the necessity
to consider a thickened hyperplane as in the limit $\sectioneps = 0$ a
typical orbit will have no points in $\Geps$.
Moreover, for a larger number of degrees of freedom this means that
the orbit returns less often to the slice and also only a small part
of phase space can be seen. Still it can be useful, if the dynamics is
dominated by a few degrees of freedom. Using \threeD{} visualization
techniques~\cite{RamVar2011} one can interactively explore the
structure and dynamics of orbits in the \psslc. Throughout this paper
we focus on \fourD{} symplectic maps and use $\sectioneps = 10^{-4}$.

For \fourD{} symplectic maps we now discuss the expected visual
appearance of the different objects in phase space: A typical chaotic
trajectory fills a \fourD{} volume in the \fourD{} phase space. In the
\psslc{} this leads to a sequence of points filling a \threeD{}
volume. A typical regular torus is a \twoD{} object embedded in the
\fourD{} phase space. In the \psslc{} this will either lead to no
points at all (i.e.\ the slice does not intersect the torus) or
typically lead to two or more \oneD{} lines. Periodic orbits usually
will not be visible in the \psslc{}. Only an appropriate choice of
\symbolsection{} will directly show periodic orbits, see
App.~\ref{sec:appl-rotat-phase2}.
To understand this reduction from the \fourD{} phase space to the
\psslc{} the analogous reduction of a \twoD{} phase space to a \oneD{}
slice is helpful: Depending on the position of the slice, regular
tori, which are \oneD{} invariant curves, will either lead to 0 (no
intersection), 1 (tangential intersection), 2 or more points. Periodic
orbits, however, will typically not be in the slice.

\begin{figure}
  \includegraphics[width=4.2cm]{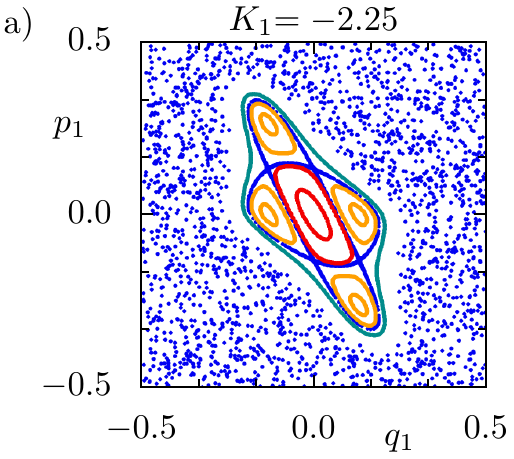}
  \includegraphics[width=4.2cm]{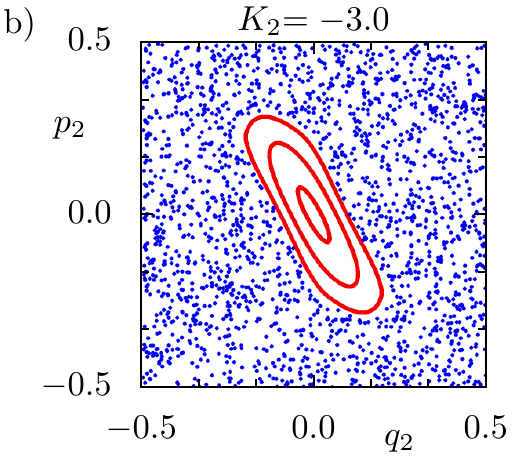}
  \caption{\label{fig:StdMaps} %
    Phase space of the uncoupled standard maps,
    Eq.~\eqref{eqn:CoupledStdMaps} with $\xic = 0$: %
    \subfiga{} $K_1=-2.25$ and $(\p_2, \q_2) = (0, 0)$.  The regular island
    (\colorregularcenter{} and \colorregularbeyondb) contains a $4:1$
    resonance (\colorresonance). Chaotic orbits (blue dots) appear
    outside of the regular island and inside a thin chaotic layer around
    the $4:1$ resonance.
    \subfigb{} $K_2=-3.0$ and $(\p_1, \q_1) = (0, 0)$. %
  }
\end{figure}

As a concrete example we consider the prototypical system
of two coupled standard maps~\cite{Fro1972} %
\begin{align}
  \label{eqn:CoupledStdMaps}
  \begin{aligned}
    \p_1' &= \p_1 + %
    \frac{K_1}{2\pi}\sin\left(2\pi \q'_1\right)  + %
    \frac{\xic}{2\pi}\sin\left(
      2\pi (\q'_1 + \q'_2)
    \right)
    \\
    \p_2' &= \p_2 + %
    \frac{K_2}{2\pi}\sin\left(2\pi \q'_2\right)  + %
    \frac{\xic}{2\pi}\sin\left(
      2\pi (\q'_1 + \q'_2)
    \right)
    \\
    \q_1' &= \q_1 + \p_1
    \\
    \q_2' &= \q_2 + \p_2
  \end{aligned}
\end{align}
where $\p_1, \p_2, \q_1, \q_2 \in [-\nicefrac{1}{2}, \nicefrac{1}{2})$
and periodic boundary conditions are imposed in each coordinate.
The parameters $K_1$ and $K_2$ control the nonlinearity of the
individual \twoD{}
standard maps in $(\p_1, \q_1)$ and $(\p_2, \q_2)$, respectively.
The parameter $\xic$ introduces a coupling between the two
degrees of freedom.
We choose as parameters of the uncoupled standard maps
$K_1 = -2.25$ and $K_2 = -3.0$ such that at each of the origins
$(\p_1, \q_1) = (0, 0)$ and $(\p_2, \q_2) = (0, 0)$
one has an elliptic fixed point, see Fig.~\ref{fig:StdMaps}.
Note that for
positive values of $K_1$ and $K_2$ the phase space would be shifted by
$\nicefrac{1}{2}$ in the $\q_1$ and $\q_2$ directions.
In both cases, Figs.~\ref{fig:StdMaps}~\subfiga{} and \subfigb,
one has a large regular region
formed by invariant tori (\colorregularcenter{} lines)
which is surrounded by a region formed by
chaotic orbits (blue dots).
For $K_1=-2.25$ the regular island contains a large
$4:1$ resonance.
Note that the regular tori form absolute barriers to the motion,
such that orbits starting in the chaotic region cannot enter
the regular island and vice versa. %
For example the blue chaotic orbit surrounding the $4:1$ resonance in
Fig.~\ref{fig:StdMaps}~\subfiga{} is not dynamically connected to the large
chaotic region also shown in blue.

We choose a strong coupling $\xic = 1.0$ to obtain a generic \fourD{}
phase space, which is far from being integrable. In this case the
origin $\fixedpoint = (\p_1, \p_2, \q_1, \q_2) = (0, 0, 0, 0)$ remains
an elliptic-elliptic fixed point which follows from %
linearization~\cite{HowMac1987, HowDul1998, LicLie92, Sko2001b}.

We use the slice condition $\p_2^*=0$, see the Appendix for a
variation of $p_2^*$ and other choices. By numerically iterating
several initial conditions until for each of them $8000$ points are
contained in the \psslc, we obtain
Fig.~\ref{fig:psslice-csm-strong-coupling}. Chaotic trajectories lead
to sequences of irregularly spread points, examples are shown as blue
dots. Regular tori, which generically are \twoD{} objects in the
\fourD{} phase space, appear as \oneD{} lines in the \psslc. They are
shown in %
\colorregularcenter, \colorregularbeyondb, \colorregularbeyonda, %
\colorregularbeyondabent, \colorresonancethreetower, %
\colorresonance, and \colorresonanceb{} %
depending on which structure in phase space they belong to. All these
structures are given by families of \twoD{} tori forming approximately
\fourD{} volumes with a \threeD{} surface towards the chaotic domain
if we neglect the Arnold web and the fractal structure of phase space.
The initial
conditions leading to the \oneD{} lines shown in
Fig.~\ref{fig:psslice-csm-strong-coupling} are chosen such that they
sample the surface of these structures in phase space evenly. This is
accomplished by choosing the initial
conditions manually based on \FLI{}
calculations~\cite{FroGuzLeg2001,LegGuzFro2003},
see Sec.~\ref{sec:fast-lyap-indic}.
By not starting orbits in the inside of the regular regions or too close
to each other one ensures a clear representation of the underlying
phase-space structures.

The regular tori in Fig.~\ref{fig:psslice-csm-strong-coupling}
appear to form a kind of
\emph{regular region} embedded in the large chaotic sea,
similar to the case of a two-dimensional map.
However, it
is important to emphasize that this regular region
is actually not a region but just a collection of regular
tori with chaotic trajectories
interspersed on arbitrarily fine scales.
One such chaotic orbit is shown as a dense cloud of blue dots.
The crucial difference between \twoD{} and \fourD{} maps is
that this blue orbit in the chaotic layer will
eventually spread into the
main chaotic sea if iterated long enough. %
As regular tori are no longer absolute barriers in phase
space~\cite{LicLie92}, the thin
chaotic zones and the main chaotic sea are connected with each other.
Thus all such chaotic trajectories close to the center will
eventually enter the outer chaotic sea.
Regular orbits of the central region are shown in \colorregularcenter{}
in Fig.~\ref{fig:psslice-csm-strong-coupling}.
Other structures given by regular tori are indicated in
different colors. %
Most of them appear twice (\colorregularbeyondb) or
four times (\colorregularbeyonda).
\label{text:3d-analogy-a}
This general feature of tori in \psslcs{} of a \fourD{} phase space
can be understood from an analogy in \threeD{} space: Here a \twoD{}
section of a \twoD{} torus will typically give two \oneD{} lines.
Furthermore, there are two groups of tori (\colorresonance{} and
\colorresonanceb) which are in the vicinity of two elliptic-elliptic
period-$7$ orbits. (This is analogous to the case of a \twoD{} map
with a resonance chain, see Fig.~\ref{fig:StdMaps}~\subfiga.) Note
that in each case only three points are close enough to the considered
\psslc{} such that their surrounding tori are visible in the slice.
The other four points are further away, see
App.~\ref{sec:vari-slice-cond}. The \colorresonance{} structure will
be investigated in detail in App.~\ref{sec:appl-rotat-phase2}.

Fig.~\ref{fig:psslice-csm-strong-coupling} together with its rotated
view~\cite{supplementalvideos} demonstrates that using a single
\psslc{} provides a lot of insight into the structures in phase space,
in particular their relative locations are revealed and sizes can be
inferred by varying the slice condition. This information is essential
to predict the number of quantum eigenfunctions that localize in the
different subregions of phase space, see
Sec.~\ref{sec:quantum-mechanics}.

\subsection{\label{sec:method-comparison}Comparison with other methods}

We now compare the \psslcs{} with other common methods, namely
\threeD{} projections of orbits, the frequency analysis, and a chaos
indicator. It turns out that they provide a complementary view of the
classical phase-space structures. However, only the \psslcs{} allow
for a visualization of quantum eigenstates in comparison with
individual regular tori, see Sec.~\ref{sec:quantum-mechanics}.
%

\subsubsection{\label{sec:orbit-projections}3D projections}

\begin{figure*}
  \includegraphics{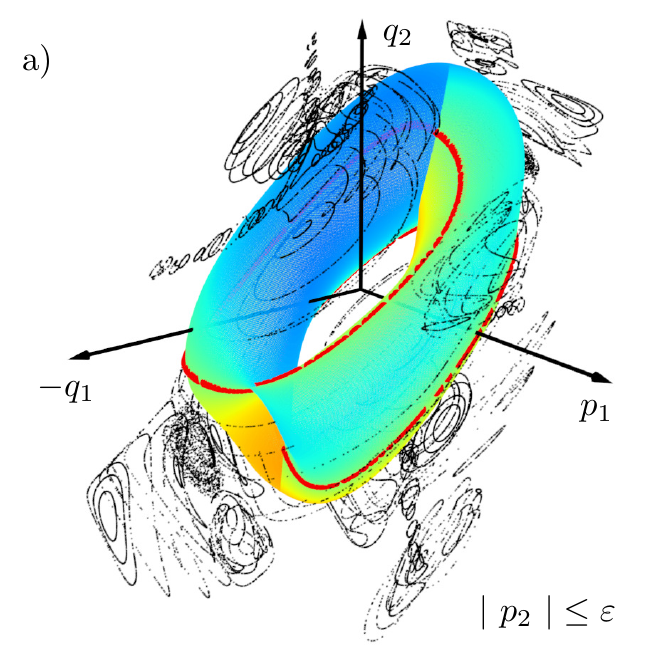}
  \includegraphics{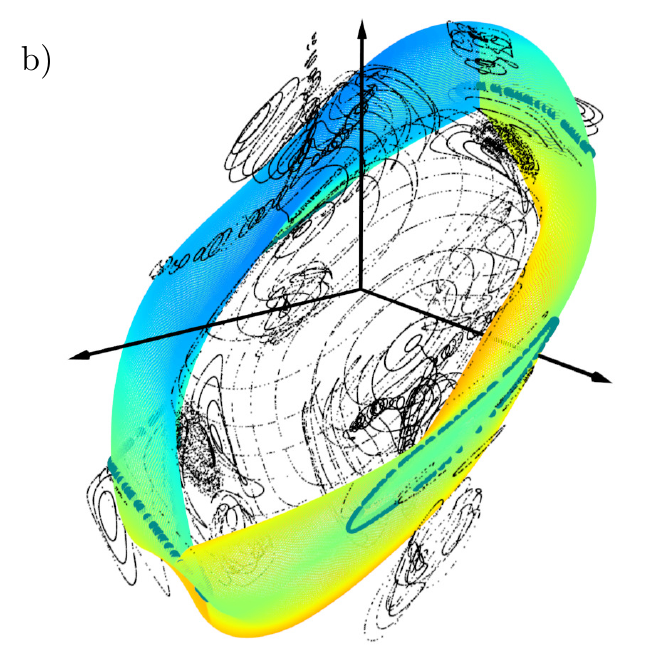}
  \\
  \includegraphics{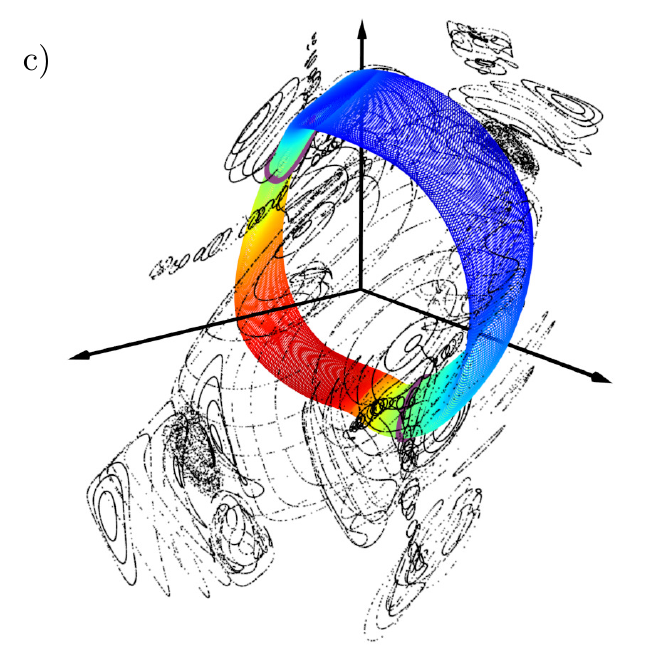}
  \includegraphics{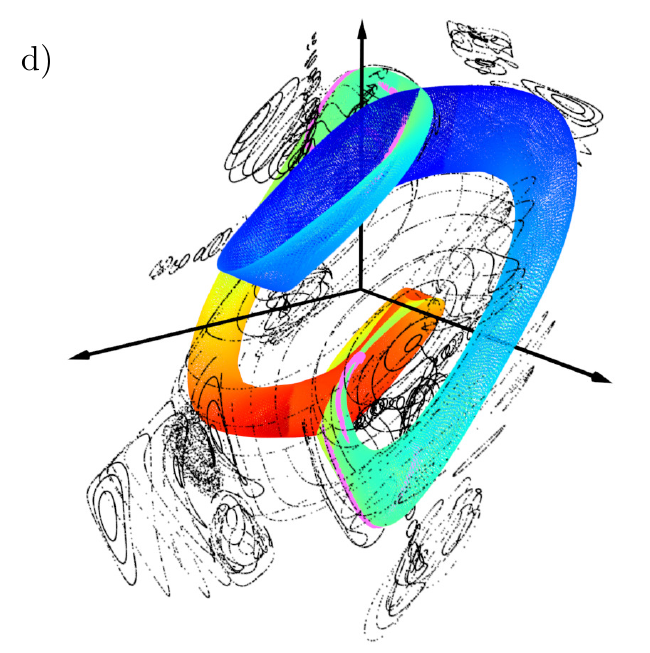}
  \\
  \includegraphics{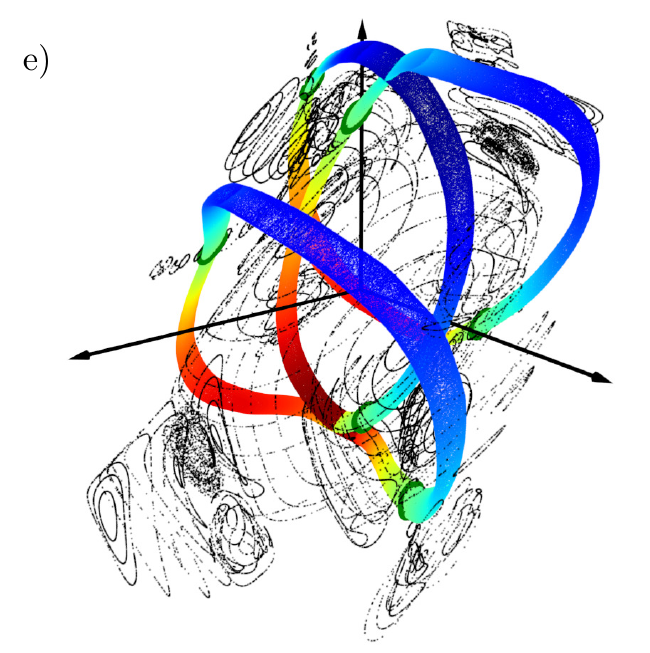}
  \includegraphics{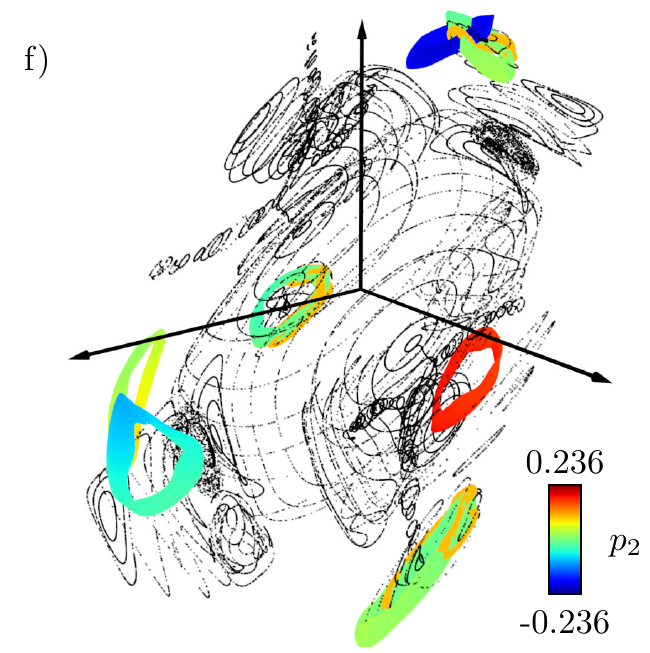}
  \caption{\label{fig:projected-orbits} %
    \Psslc{} for $|\p_2| \le \sectioneps$
    of two coupled standard maps
    with \threeD{} projections for different regular tori from
    Fig.~\ref{fig:psslice-csm-strong-coupling}: %
    \subfiga{} \colorregularcenter{}, 
    \subfigb{} \colorregularbeyondb{}, 
    \subfigc{} \colorregularbeyonda{}, 
    \subfigd{} \colorregularbeyondabent{}, 
    \subfige{} \colorresonancethreetower{}, and 
    \subfigf{} \colorresonance{}. 
    For the projection the $\p_2$ coordinate is encoded in color
    corresponding to the colorbar. %
    \MOVIEREF }
\end{figure*}

Using \psslcs{} provides a global view of the geometry of the
underlying \fourD{} phase space by displaying several objects in the
same plot. Alternatively, one can concentrate on one or few orbits and
display all points by a \threeD{} projection~\cite{Fro1970a}
and encode the value of the projected coordinate by a color scale, see
Fig.~\ref{fig:projected-orbits}. This approach has been called
\emph{method of color and rotation}, see Refs.~\cite{PatZac1994,
  KatPat2011, EftHar2013p}.
The plots of the orbit projections reveal the underlying
topology, i.e. which structures of the \psslc{} \Geps{} (black)
are connected.
Figs.~\ref{fig:projected-orbits}~\subfiga~--~\subfigc{}
show projected orbits belonging to
the \colorregularcenter{},
\colorregularbeyondb{}, and \colorregularbeyonda{} region from
Fig.~\ref{fig:psslice-csm-strong-coupling}.
The projection shows that they indeed share
the same topology. A closer investigation reveals that
the torus (\colorregularbeyondabent)
shown in Fig.~\ref{fig:projected-orbits}~\subfigd{}
is also of the same type
and just bent in a more complicated way. The highlighted tori shown in
Figs.~\ref{fig:projected-orbits}~\subfiga~--~\subfigd{} have been
called \emph{rotational}~\cite{VraBouKol1996,KatPat2011}.

Fig.~\ref{fig:projected-orbits}~\subfige{} shows the projection of an
orbit from the \colorresonancethreetower{} structure around the
central regular region. Such tori have been called \emph{tube
  tori}~\cite{VraBouKol1996,KatPat2011}. %
A frequency analysis, see Sec.~\ref{sec:frequencies}, shows that the
orbits of the \colorresonancethreetower{} structure belong to the
stable vicinity of the elliptic $-1:3:0$ rank-$1$
resonance~\cite{Tod1994}. Hence, we conclude that the tube tori result
from coupled rank-$1$ resonances.

Finally, Fig.~\ref{fig:projected-orbits}~\subfigf{} shows the
projection of an \colorresonance{} orbit from the elliptic-elliptic
vicinity of the period-$7$ orbit of
Fig.~\ref{fig:psslice-csm-strong-coupling}. Seven parts can be seen in
the projection at the same time, while just three fulfill the slice
condition of Fig.~\ref{fig:psslice-csm-strong-coupling}.

Only one or a few orbits can be displayed simultaneously using
\threeD{} projections, while many can be visualized in the \psslc{}.
Hence, using \psslcs{} and projections of orbits in combination is
very instructive for the understanding of higher-dimensional maps.

\subsubsection{\label{sec:frequencies}Frequency analysis}

\begin{figure}
   \includegraphics{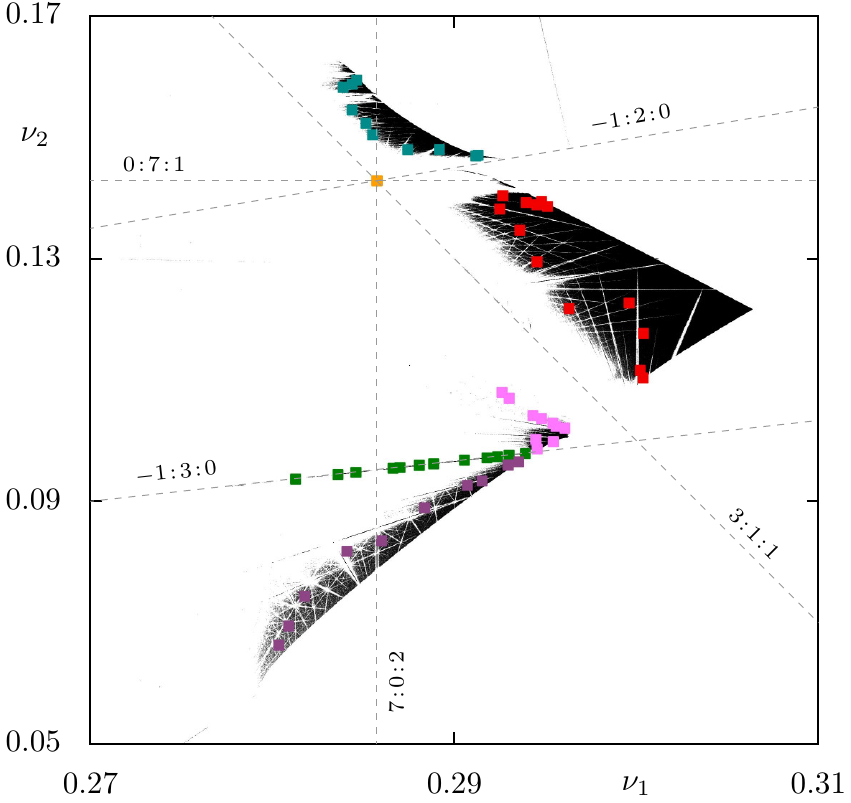}
   \caption{\label{fig:frequencies} Frequency plane of two coupled
     standard maps at strong coupling $\xic = 1.0$. The colored
     squares correspond to the colored regular tori of
     Fig.~\ref{fig:psslice-csm-strong-coupling}. The black points
     correspond to additional regular tori.}
\end{figure}

We now compare the \psslcs{} with the frequency
analysis~\cite{MarDavEzr1987, Las1993, ChaWigUze2003, SetKes2012}.
This method associates each regular torus with its two fundamental
frequencies $(\nu_1, \nu_2) \in [0,1[^2$, which are displayed in the \twoD{}
frequency plane, see Fig.~\ref{fig:frequencies}.
In the plot the black points represent regular tori obtained by
starting $10^8$ initial conditions (with randomly chosen
$p_1, p_2 \in [-0.2, 0.2]$ and $q_1, q_2 \in [-0.2, 0.2]$) %
in the \fourD{} phase-space. Note that for strongly coupled maps far
from integrability a sampling on \twoD{} planes is not sufficient as
this will typically miss some regions with regular motion. Each
frequency pair is calculated from $N=4096$ iterations using a fast,
analytical interpolation method (Sec.~4.2.4 in
Ref.~\cite{BarBazGioScaTod1996}). The error of this method scales with
$N^{-4}$ like the original method of Laskar~\cite{Las1993}. To decide
whether an orbit is regular we use the frequency criterion
\begin{align}
  \label{eq:frequency-criterion}
  \max{\left(|\nu_1 - \tilde{\nu}_1|,|\nu_2 - \tilde{\nu}_2|\right)} < 10^{-7},
\end{align}
where the frequency pair $(\tilde{\nu}_1, \tilde{\nu}_2)$ is calculated from $N$
further iterations. Note that the frequencies $(\nu_1, \nu_2)$ are
only defined up to a unimodular transformation (Sec.~15 in
Ref.~\cite{Bor1927} and Refs.~\cite{DulMei2003,GekMaiBarUze2007}).
Using the \psslcs{} we can choose the frequencies consistently such
that regular tori which are close in phase space are also close in the
frequency plane. Explicitly, we make the following transformations: i)
if $\nu_i>0.5$ then $\nu_i \mapsto 1-\nu_i$, ii) if $\nu_2>\nu_1$ then
$(\nu_1,\nu_2) \mapsto (\nu_2,\nu_1)$, iii) for tori of the type shown
in Fig.~\ref{fig:projected-orbits}~\subfigd{} with $\nu_2>0.25$ we use
$(\nu_1, \nu_2) \mapsto (\nu_1, -4 \nu_1 + \nu_2)$. The resulting
frequency pairs reside in a small region of the frequency plane, see
Fig.~\ref{fig:frequencies}.

Additionally, the frequencies of the regular tori shown in
Fig.~\ref{fig:psslice-csm-strong-coupling} are displayed as squares
with the corresponding color. This provides a connection between
phase-space structures and areas in the frequency plane: E.g., the
\colorregularcenter{} points indicate that the area to the right
corresponds to the central regular region. In fact the linearization
of the central elliptic-elliptic fixed point \fixedpoint{} gives
frequencies $(\nu_1, \nu_2) = (0.30632, 0.12173)$, which coincide with
the rightmost tip in the frequency plane. The sharp edges emanating
from the tip correspond to elliptic \oneD{} tori, i.e., the limiting
case of \twoD{} tori for which one action becomes
zero~\cite{LanRicOnkBaeKet2013unpubl}.

The frequency plane is organized by rank-$1$ resonance lines, on which
the frequencies fulfill~\cite{Las1993}
\begin{align}
\label{eq:resonance-driven}
m_1 \cdot \nu_1 + m_2 \cdot \nu_2 = n
\end{align}
with $m_1, m_2, n$ being integers. The most important ones are
displayed in Fig.~\ref{fig:frequencies} abbreviating
Eq.~\eqref{eq:resonance-driven} by $m_1:m_2:n$. We stress that the
resonance lines lead to gaps in the \fourD{} phase space, clearly
visible in the \psslcs{}, see
Fig.~\ref{fig:psslice-csm-strong-coupling}. E.g., the $-1:2:0$
resonance separates the \colorregularcenter{} and the
\colorregularbeyondb{} region and the $3:1:1$ resonance separates the
\colorregularcenter{} and the \colorregularbeyondabent{} region. The
tori of the type shown in Fig.~\ref{fig:projected-orbits}~\subfigd{}
(\colorresonancethreetower{} in
Fig.~\ref{fig:psslice-csm-strong-coupling}) demonstrate an advantage
of the \psslcs{}: Performing a frequency analysis for these tori gives
frequency pairs far away from the frequency region shown in
Fig.~\ref{fig:frequencies}. If one ignores a fundamental frequency
with small amplitude and instead chooses two frequencies with large
amplitude, the frequency pairs reside within the frequency region
displayed in Fig.~\ref{fig:frequencies}. However, they collapse on the
resonance line $-1:3:0$. In contrast, the representation of these tori
in the \psslc{} highlights both their relative location to other tori
and that they are a two-parameter family of tori. The same is also
true for the tori in the vicinity of the elliptic-elliptic period-$7$
orbits (\colorresonance, \colorresonanceb{} in
Fig.~\ref{fig:psslice-csm-strong-coupling}). Choosing two frequencies
with large amplitude in this case results in all frequency pairs
collapsing to the point $(\nu_1, \nu_2) = (\nicefrac{2}{7},
\nicefrac{1}{7})$.

These results demonstrate how the frequency plane with its resonance
lines relates to the structures of the \fourD{} phase space. The
\psslcs{} reveal the topology and geometrical relevance of the
resonance gaps. In particular, the \psslcs{} help to consistently
assign frequencies to the regular tori in systems that are far away
from integrability.

\subsubsection{\label{sec:fast-lyap-indic}Fast Lyapunov Indicator}

\begin{figure}
  \includegraphics{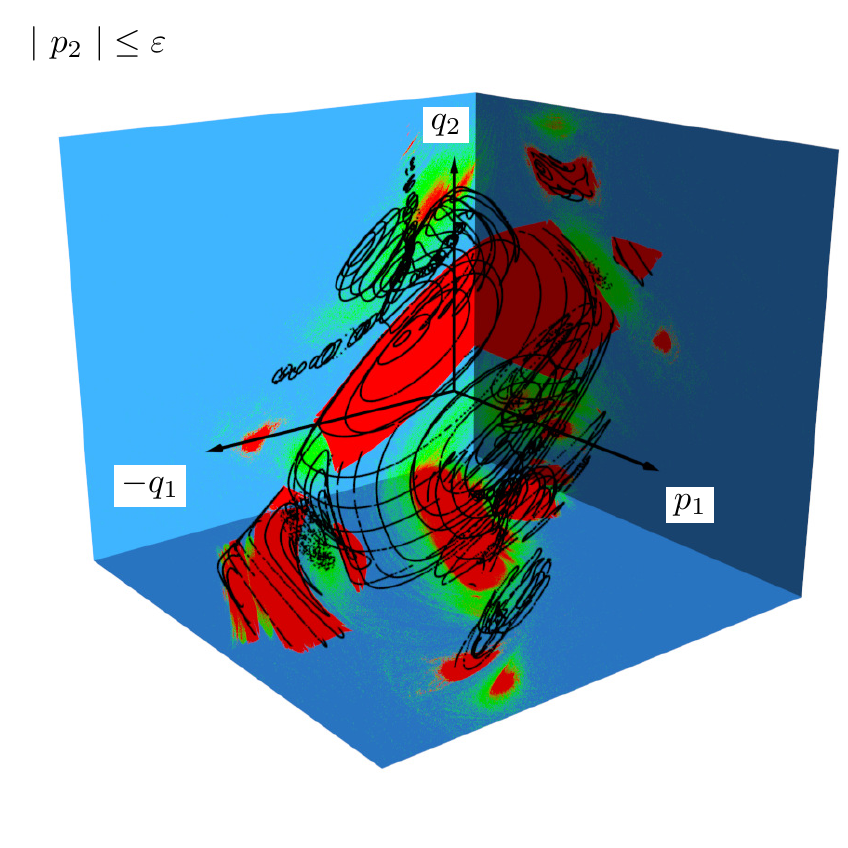}
  \caption{\label{fig:fli-2d-plane-in-3d} \Psslc{} for $|\p_2|
    \le \sectioneps$ of two coupled standard maps at strong
    coupling $\xic = 1.0$ with three \twoD{} \FLI{} planes. The
    colormap is chosen such that points with $\FLI{} < 30$ are shown in
    red, points with $\FLI{} > 150$ are shown in blue, and intermediate
    values are marked green.
    The three planes are chosen away from the central
    elliptic-elliptic fixed point for better
    visibility.
    \MOVIEREF
    }
\end{figure}

We now compare the \psslcs{} with one example of a chaos indicator.
Such chaos indicators associate with each initial condition a value
describing the chaoticity of the corresponding
trajectory~\cite{Sko2001, MafDarCinGio2011, Zot2012}.
We consider as an example the Fast Lyapunov Indicator
(\FLI)~\cite{FroLegGon1997, FroGuzLeg2000, FroLeg2000},
see Fig.~\ref{fig:fli-2d-plane-in-3d}.
To interpret the results of this indicator we calculate a
histogram for a set of initial
conditions in the \fourD{} phase space. Small values correspond to
regular motion and large values correspond to chaotic motion which
allows for adjusting the colorscale, see
Fig.~\ref{fig:fli-2d-plane-in-3d}.
In this plot the \FLI{} is displayed on three mutually perpendicular
planes. These are placed such that most of the regular region is
visible.
The \FLI{} is calculated using $4096$ iterations averaging over the last
$50$ values. The initial tangent vector is $(1, 1, 0.5 \cdot ( \sqrt{5}
- 1), 1)$ inspired by Ref.~\cite{LegGuzFro2003}.
The colorscale is chosen such that red regions coincide with regions
of regular motion as can be seen from the orbits (black) in the
\psslc. Regions with intermediate \FLI{} values are shown in green.
Large \FLI{} values correspond to strongly chaotic motion and are shown
in blue.

Such a \FLI{} representation is also useful for selecting regular
orbits for the \psslc. For Fig.~\ref{fig:psslice-csm-strong-coupling}
orbits from the border of the regular regions are used. Including also
orbits further inside would lead to a less clear visualization.

Another application of the \FLI{} is the estimation of the size of the
regular region. For this one starts trajectories on a grid in the
\fourD{} phase space and determines the fraction of orbits with a
small \FLI{} value, i.e.\ corresponding to the red regions of
Fig.~\ref{fig:fli-2d-plane-in-3d}. Using $128 ^ 4$ initial points and
a maximum \FLI{} value of $30$ we find the approximate regular
phase-space fraction to be $1.4 \cdot 10^{-3}$. Quantum mechanically
this gives an estimate of the number of regular eigenstates, see
Sec.~\ref{sec:husimi-slices}.

Fig.~\ref{fig:fli-2d-plane-in-3d} shows that the plot for the regular
tori in the \psslcs{} complements the plot of chaos indicators like
the \FLI{}. An advantage of \psslcs{} is that the geometry of
individual tori is visible, which is particularly important when
considering the quantized map, for which regular eigenstates
concentrate on such tori.

\section{\label{sec:quantum-mechanics}Structure of quantum states}

By means of the \psslcs{} we are able to relate the quantum mechanical
properties of higher-dimensional systems with the underlying classical
structures. For this we first introduce the corresponding quantum map
and the computation of eigenstates. Then the Husimi representation of
eigenstates is visualized on the \psslcs{} to investigate the
semi-classical eigenfunction hypothesis.

\subsection{Quantized map}\label{sec:quantized-map}

The classical map~\eqref{eqn:CoupledStdMaps}
arises from the stroboscopic view of a kicked Hamiltonian of the form
\begin{align}
  \label{eq:Hamiltonian-periodic-driving}
  H(\vec\p, \vec\q\,) = T(\vec\p\,) + V(\vec\q\,)
  \sum\limits_{n \in \Z} \delta( t - n )
\end{align}
where the period of the driving is $1$.
Quantum mechanically the time evolution
\begin{align}
  \label{eq:qm-time-evolution-step}
  \ket{\psi (t + 1)} = \Uop \ket{\psi (t)},
\end{align}
of a state $ \ket{\psi (t)}$ after one time period
is fully determined by the unitary operator \Uop.
The usual quantization for
\twoD{} maps, see
e.g.\ Refs.~\cite{BerBalTabVor79, ChaShi86, KeaMezRob1999,
  DegGra2003b},
straightforwardly carries over to the \fourD{} case, see
e.g.\ Refs.~\cite{RivSarAlm2000, Lak2001}.
Explicitly we have
\begin{align}
  \Uop = \exp\left( -\ioverhbar
    V( \hat{\vec{\q}}\, ) \right)
  \exp\left( -\ioverhbar
    T( \hat{\vec{\p}}\, ) \right)
\label{propagator}
\end{align}
where %
$V(\vec{\q}\,) = %
\frac{K_1}{4\pi^2}\cos\left(2\pi \q_1\right) + %
\frac{K_2}{4\pi^2}\cos\left(2\pi \q_2\right) + %
\frac{\xic}{4\pi^2}\cos\left(2\pi (\q_1 + \q_2) \right)$ %
and %
$T(\vec{\p}\,) = %
\frac{1}{2} \left( \p_1^2 + \p_2^2 \right) $.
Here $\heff = 2\pi \hbareff$ is the effective Planck's constant which
is Planck's constant $h$ divided by the $\Ndof$-th
root of the size of one unit cell \volumeunitcellfD{}
in phase space,
\begin{align}
  \heff = \frac{h}{
    \left(\volumeunitcellfD\right)^{\nicefrac{1}{\Ndof}}}
\end{align}
where in our case we have the number of degrees of freedom $\Ndof = 2$.

Of particular interest are the stationary states, i.e.\
the eigenstates $\ket{\psiqnums{j}}$ of \Uop, defined by
\begin{align} \label{eq:eigenstates}
  \Uop \ket{\psiqnums{j}} = \ue^{\ui \varphi_j} \ket{\psiqnums{j}}
\end{align}
with the eigenphases $\eigenphase_j$.
Due to the periodicity of the classical phase space
in the $\p_i$ directions
one can express
the time evolution operator \Uop{} in a discrete position basis
$\ket{\varqn} \equiv \ket{\vec{\q}_{n_1, n_2}}$ with $0 \leq n_i \leq
\QMdimperdof -1$ where $\QMdimperdof$ is the number of grid points in
each direction.
This gives rise to a $\QMdim$-dimensional Hilbert space and the quantization
condition
\begin{align}
  \label{eq:qm-condition-hbar}
  \heff = \frac{1}{\QMdimperdof}.
\end{align}
With this the grid in position space reads
\begin{align}
  \label{eq:qm-grid-q-pre}
  \varqn[n] = %
  \vec{\q}_0 + \frac{1}{\QMdimperdof} \vec{n}
\end{align}
with $\vec{\q}_0 = (-\nicefrac{1}{2}, -\nicefrac{1}{2})$.

Now we can express the propagator \Uop{} by a
finite $\QMdim \times \QMdim$ unitary matrix (for even $N$)
\begin{align}
  \Uopsymb_{\vec{n} \vec{k}} & {} \equiv
  \bra{\varqn} \Uop \ket{\varqn[k]} \notag
  \\
  \label{eq:U-position}
  & {} =
  \heff[2] \
  \ue^{-\ioverhbar V(\varqn)}
  \sum\limits_{j_1=0}^{\QMdimperdof - 1}
  \sum\limits_{j_2=0}^{\QMdimperdof - 1}
  \ue^{-\ioverhbar \left(
      T(\varpj)
      + \varpj \left(\varqn - \varqn[k]\right)
    \right)},
\end{align}
where $\varpj = \vec{\p}_0 + \frac{1}{N} \vec{j}$. Finding the
solution of \eqref{eq:eigenstates}, i.e.\ the eigenphases and
eigenstates of the system, therefore reduces to the numerical
diagonalization of the unitary matrix~\eqref{eq:U-position}.
The matrix size scales like $\heff[-2]$, in contrast to \twoD{} maps
where it scales like $\heff[-1]$.
This shows that numerical studies of higher-dimensional systems in the
semi-classical limit of small $\heff$ require much more computational
effort.

\subsection{Lanczos algorithm}
\begin{figure*}
  \begin{tabular}{ll}
    \includegraphics{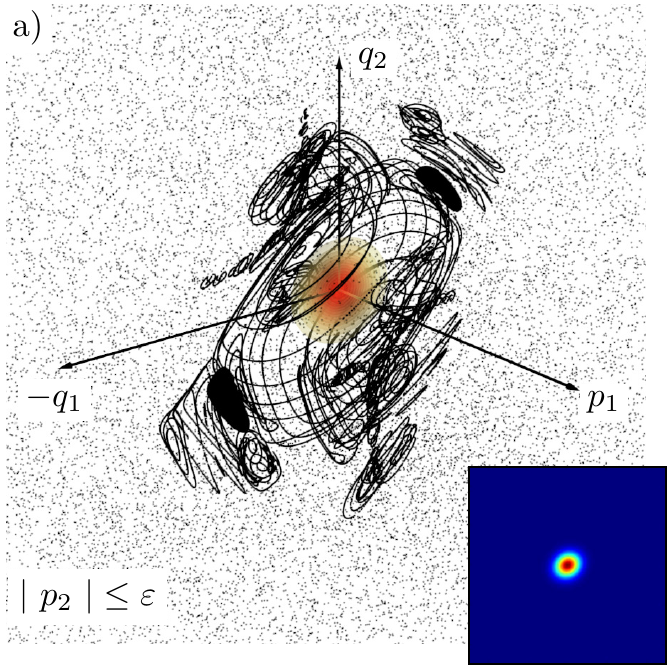}
    &
    \includegraphics{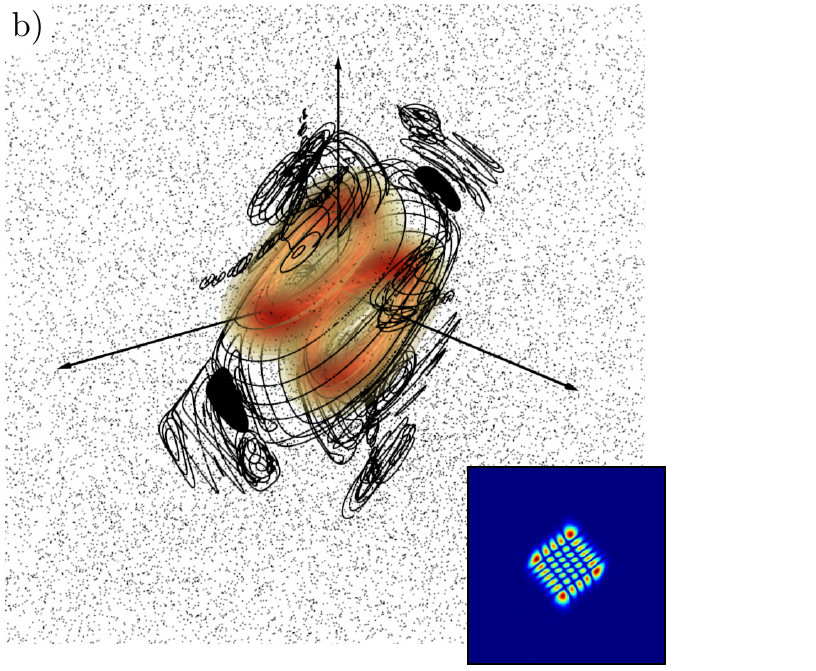}
    \\
    \includegraphics{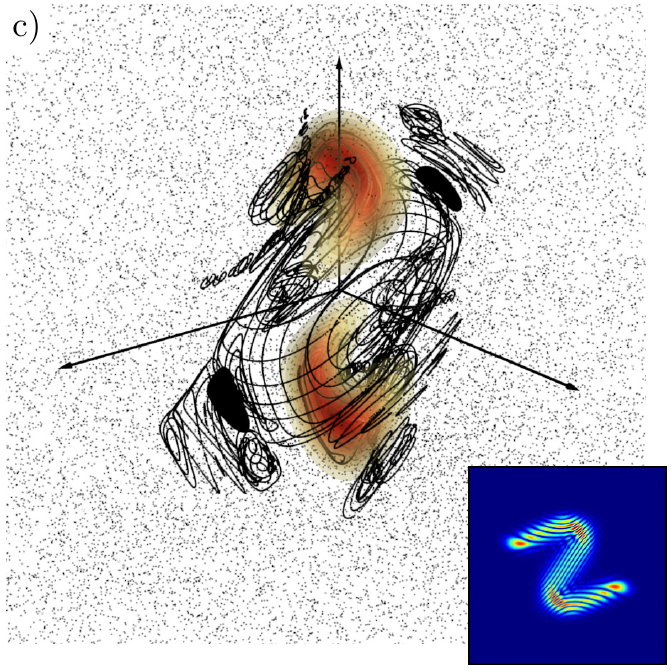}
    &
    \includegraphics{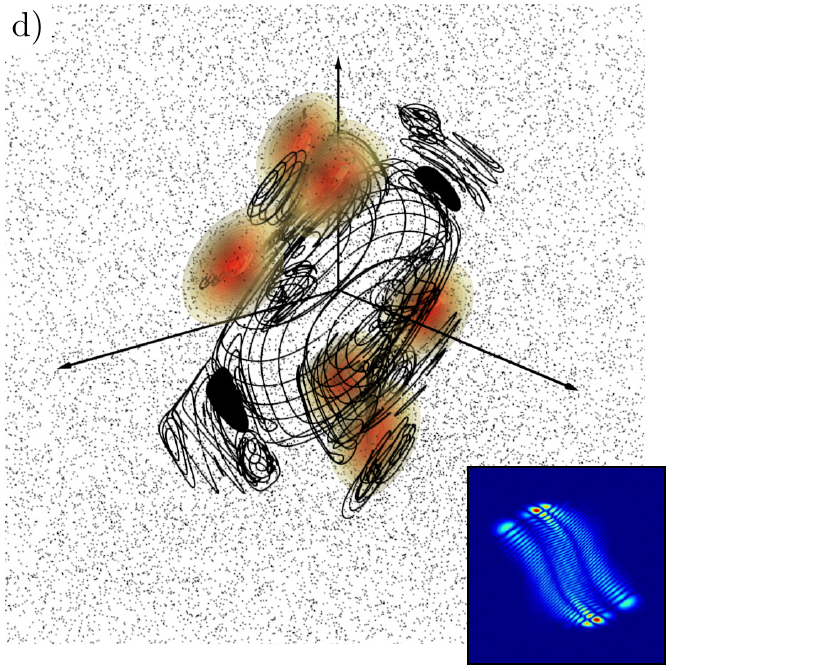}
    \\
    \includegraphics{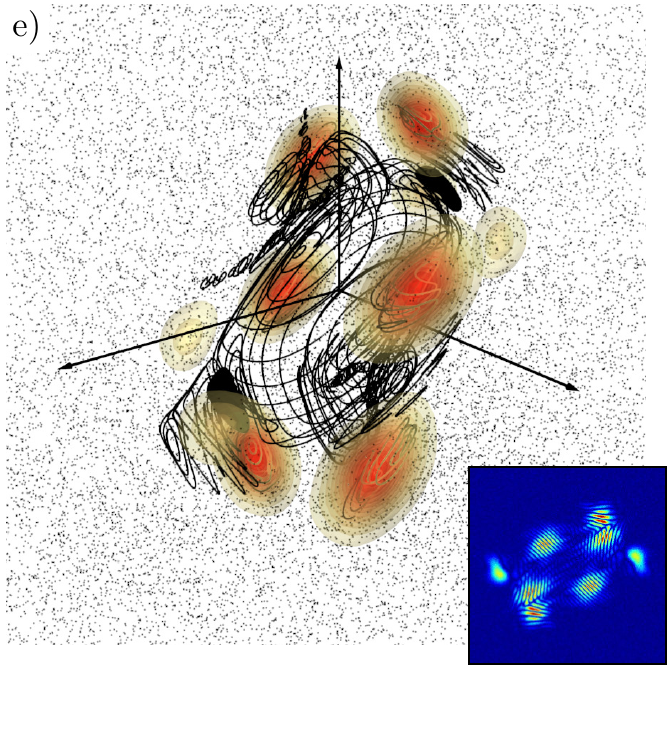}
    &
    \includegraphics{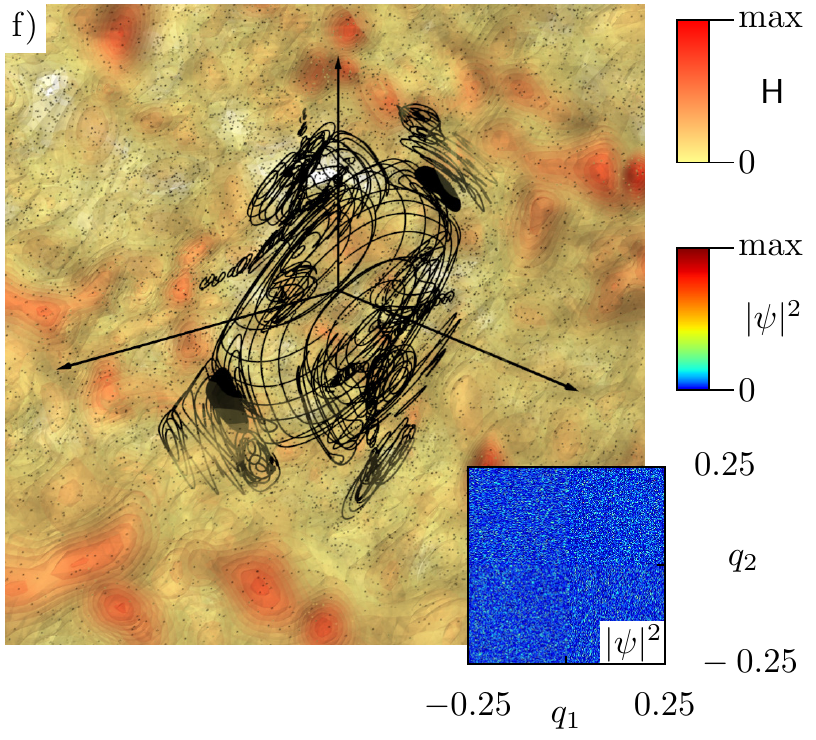}
  \end{tabular}
  \caption{%
    Husimi representation \Husimi{} of eigenstates of the
    time-evolution operator \Uop{} for $\heff = \frac{1}{500}$ in the \psslc{}
    and position-space probability-density $|\psi|^2$ (insets).
    \subfiga{} ground state $\ket{\psiqnums{0, 0}}$
    of the central regular region.
    \subfigb{} excited state $\ket{\psiqnums{4, 6}}$.
    \subfigc{} state concentrated on \colorregularbeyondabent{}
    structure shown in Fig.~\ref{fig:psslice-csm-strong-coupling}.
    \subfigd{} state concentrated on \colorresonancethreetower{}
    structure shown in Fig.~\ref{fig:psslice-csm-strong-coupling}.
    \subfige{} state concentrated on vicinity of period-$7$ orbit
    (\colorresonance{} and \colorresonanceb{} structure in
    Fig.~\ref{fig:psslice-csm-strong-coupling}).
    \subfigf{} chaotic eigenstate where the octant in the front
    has been removed for better visibility.
    \MOVIEREF
   }
  \label{fig:husimi-gallery}
\end{figure*}

For the semi-classical behavior of
eigenstates,
i.e.\ in the limit of small $\heff$,
we have to consider large matrices.
This makes
it necessary to use appropriate diagonalization schemes which allow
for computing a few  rather than all eigenstates.
One possibility is the Lanczos algorithm~\cite{CulWil1985}
which transforms
$\Uopsymb$ into tridiagonal form.
The method is particularly efficient if the
application of the unitary operator to a vector can
be performed fast, in our case by a fast Fourier-transform
 as in Ref.~\cite{KetKruGei1999}.
This leads to a computational effort of $N^4 \ln{N}$
instead of $N^6$ for the direct diagonalization.

In order to use this approach for the computation of eigenstates
concentrated on a specific region in the classical phase space, the
initial vector of the Lanczos algorithm
has to be chosen appropriately.
Usually a random vector is used as the initial state
$\ket{\Lanczosinitial}$. Instead we choose a coherent state
$\ket{\Lanczosinitial} = \ket{\coherent(\vec\p_0, \vec\q_0)}$
\begin{align}
  \label{eq:qm-coherent-state-definition-q}
  \braket{\vec{\q}}{\coherent(\vec\p_0, \vec\q_0)} =
  \frac{1}{\sqrt{\pi\hbareff}} \exp\left(
    -\frac{(\vec{\q} - \vec{\q}_0)^2}{2\hbareff}
    + \ui \frac{ {\vec{\p}_0}\vec{\q} }{\hbareff}
  \right)
\end{align}
concentrated at a point $(\vec\p_0, \vec\q_0)$.
By this it is possible to selectively calculate those eigenstates
which have a large overlap with $\ket{\Lanczosinitial}$:
After the transformation to tridiagonal form we
sequentially~\footnote{ %
  This sequential computation relies on routines which calculate part
  of the spectrum and the corresponding eigenvectors
  $\ket{\psiqnums{j}}$. Here we used \texttt{dstevr} of a recent
  \ATLAS{} implementation.} calculate all eigenvectors
$\ket{\psiqnums{j}}$. %
By construction of the Lanczos algorithm the first component of each
eigenvector gives the overlap with the initial coherent state
$\braket{\psiqnums{j}}{\Lanczosinitial}$.
We select the states with largest overlap
and transform them back to position space representation.
Finally, one has to verify whether the resulting state
really is an eigenstate as the Lanczos algorithm
introduces spurious solutions due to round-off errors.
As a criterion the residuum
\begin{align}
  \label{eq:residuum-quantum-states}
  \left\| \left( \Uop - \ue^{\ui \varphi_j} \right)
    \ket{\psiqnums{j}} \right\|^2,
\end{align}
is computed and we only consider states having a value smaller than
$10^{-10}$. Note that for weakly coupled systems the initial state for
the Lanczos algorithm could, instead of a coherent state, also be a
direct product of eigenstates of the uncoupled \twoD{} systems.

\subsection{Husimi representation of
  eigenstates}\label{sec:husimi-slices}

While the eigenstates of a \fourD{} map can still
be visualized in position representation,
this does not allow for understanding where
the eigenstates concentrate in phase space.
One possible phase-space representation of a state
is the Husimi representation
(see Refs.~\cite{Hus1940, NonVor1998, Bae2003}
and references therein)
\begin{align}
  \Husimi_{\psiqnums{}}(\vec\p, \vec\q\,) = %
  \frac{1}{\heff[2]} | \braket{\coherent(\vec\p, \vec\q\,)}{\psi} | ^ 2
\end{align}
which is the projection of the state $\ket{\psi}$ onto coherent states
$\coherent(\vec\p_0, \vec\q_0)$,
where also the periodicity of the phase space has to be taken into
account by periodizing the coherent states.

To visualize the Husimi function on the \psslc, i.e.\
$\Husimi_{\psiqnums{}}(\p_1, \p_2 = 0, \q_1, \q_2)$,
semi-transparent iso-surfaces are used
with a color association such that red corresponds
to high intensity, see Fig.~\ref{fig:husimi-gallery}.
While classical orbits are displayed in a \psslc{} $\Gamma_\varepsilon$,
the Husimi function is computed on the hyperplane $\Gamma$.
As the fraction of the volume of the regular phase-space region can be
estimated to be $1.4 \cdot 10^{-3}$, we expect for $\heff = \frac{1}{500}$ approximately $350$
regular and $249\,650$ chaotic eigenstates.

Fig.~\ref{fig:husimi-gallery}~\subfiga{} shows the ``ground state'' of
the central regular region (\colorregularcenter{} in
Fig.~\ref{fig:psslice-csm-strong-coupling})
which concentrates around the elliptic-elliptic fixed point.
It is approximately given by
a Gaussian shape in the Husimi function $\Husimi_{\psiqnums{}}$.
In the position-space
probability-density $|\psiqnums{} (\vec{\q}\,)| ^2 $ it is also
approximately given by a Gaussian distribution.
Fig.~\ref{fig:husimi-gallery}~\subfigb{} shows an excited state
of the central region. Its Husimi function nicely shows that this state
predominantly lives on the classical torus which corresponds to two
\oneD{} lines in the \psslc{}. The quantum
numbers $(4, 6)$ of this state can be read off from the nodal lines
in the position-space probability-density.

Figs.~\ref{fig:husimi-gallery}~\subfigc{} to~\subfige{}
show eigenstates concentrating on more complicated phase-space structures,
corresponding to the regular tori shown in
Fig.~\ref{fig:psslice-csm-strong-coupling} (\colorregularbeyondabent,
\colorresonancethreetower, and \colorresonance{}).  Note
that the eigenstate concentrating on the period-$7$ orbit extends over
both chains (\colorresonance{} and \colorresonanceb{} in
Fig.~\ref{fig:psslice-csm-strong-coupling}). This is also the reason,
why the position-space probability-density plot shows eight maxima:
These are given by two period-$7$ orbits each of which shows $4$ distinct points
in the projection onto the $(\q_1, \q_2)$ plane.
Finally, Fig.~\ref{fig:husimi-gallery}~\subfigf{} shows a chaotic
state which is concentrated on the chaotic sea and decays rapidly into
the regular region. Looking at the eigenstates on a logarithmic
representation reveals tunneling tails of chaotic states into the
regular region and of regular states into the chaotic sea (not shown).

Fig.~\ref{fig:husimi-gallery} clearly shows that the eigenstates
either concentrate on regular tori or within the chaotic sea providing
visual confirmation of the semi-classical eigenfunction hypothesis for
\fourD{} maps. This is made possible by the use of \psslcs{}.

\section{\label{sec:conclusion}Summary and outlook}

To understand the dynamics of \fourD{} symplectic maps, we use
\psslcs{}. They provide the basis for a similar level of understanding
as for \twoD{} maps. Such a visualization is of particular importance
for generic systems being far away from integrability. Classically,
the \psslcs{} reveal a regular region embedded in a large chaotic sea
for the prototypical example of two coupled standard maps, see
Fig.~\ref{fig:psslice-csm-strong-coupling}. The regular region
consists of several substructures of different topology, which are
separated by gaps. The geometrical relation of these substructures in
phase space enables to consistently assign frequencies to the regular
tori and helps to interpret the results of the frequency analysis. We
identify resonance gaps by a combination of \psslcs{} and frequency
analysis. Moreover, we conclude that the so-called tube tori result
from coupled rank-$1$ resonances.

A comparison of the \psslcs{} with orbit projections and a chaos
indicator shows that they nicely complement each other. However, an
important motivation for the use of \psslcs{} is the investigation of
quantum mechanical properties in higher-dimensional systems. To relate
the structure of eigenstates with corresponding classical structures,
the comparison has to be done in phase space. The \psslcs{} are best
suited for this purpose, since they visualize both several individual
orbits at the same time and the global geometry in phase space. We
display several eigenstates of the time-evolution operator in the
Husimi representation on the \psslc{} together with classical orbits.
By this we can confirm the semi-classical eigenfunction hypothesis,
that states either concentrate on regular tori or in the chaotic
region. Based on this an investigation of dynamical tunneling between
a regular and the chaotic region is possible.

The \psslcs{} also allow for displaying families of elliptic \oneD{}
tori that provide the skeleton around which the \twoD{} tori are
organized~\cite{Lan2012,LanRicOnkBaeKet2013unpubl}. In the future, the
\psslcs{} may be used to display invariant phase-space structures like
\oneD{} tori or stable and unstable manifolds of fixed points and
periodic orbits. Another important application of \psslcs{} is the
determination of those phase-space structures at which trapping of
chaotic orbits occurs in systems far away from
integrability~\cite{Lan2012}. These long-trapped orbits are the key to
the understanding of power-law recurrence-time statistics.

We hope that the method of \psslcs{} will be useful for the
understanding of the phase-space structure and the dynamics of
\fourD{} maps and possibly higher-dimensional systems.

\begin{acknowledgments}
We thank Srihari Keshavamurthy and Peter Schlagheck
for stimulating discussions. We further acknowledge discussions with
Jacques Laskar, Haris Skokos and Matthaios Katsanikas.
We thank the Center for Information Services and High Performance Computing
(\textsc{zih} Dresden) for access to the computing facilities.
Furthermore, we acknowledge support by the Deutsche Forschungsgemeinschaft
within the Forschergruppe 760 ``Scattering Systems with Complex Dynamics.''
All \threeD{} visualizations were created using
\mayavi~\cite{RamVar2011}.
Finally, we would like to thank
{\it Back to the Future Part III}
for the quote~\cite{BTTFIII}:
{\it You're just not thinking fourth dimensionally!} --
{\it Right, right. I have a real problem with that.}

\end{acknowledgments}

\appendix*   

\section{\label{sec:alt-phase-space}General 3D phase-space slices}

As only a part of phase space is visible
in a \psslc{} it is instructive to consider other \psslcs{},
e.g., by shifting the slice condition, choosing other coordinates for
the slice, or by using adapted non-orthogonal coordinates. This is
illustrated in the following sections.

\subsection{\label{sec:vari-slice-cond}Shifting the slice condition}

\begin{figure*}
  \includegraphics{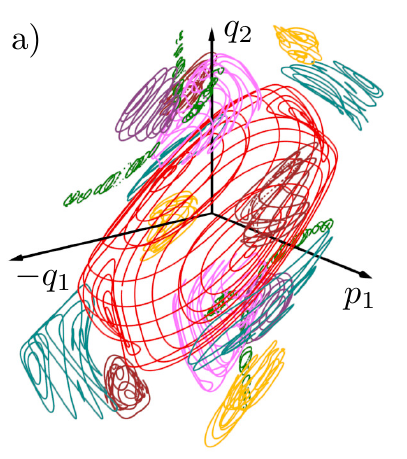}
  \includegraphics{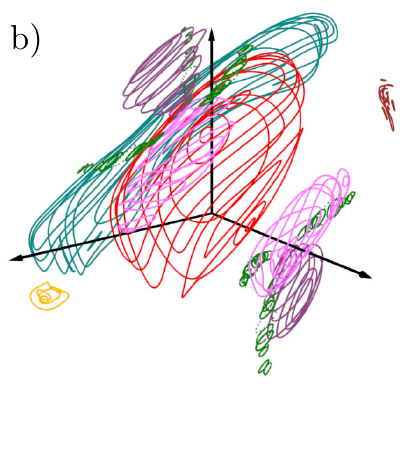}
  \includegraphics{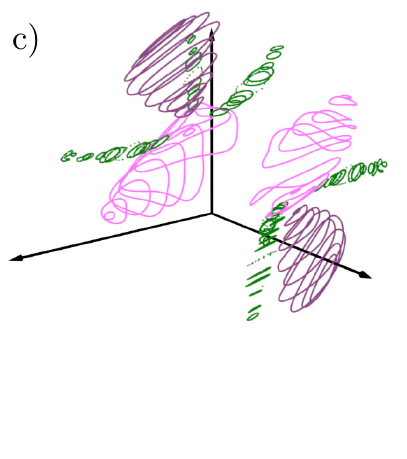}
  \includegraphics{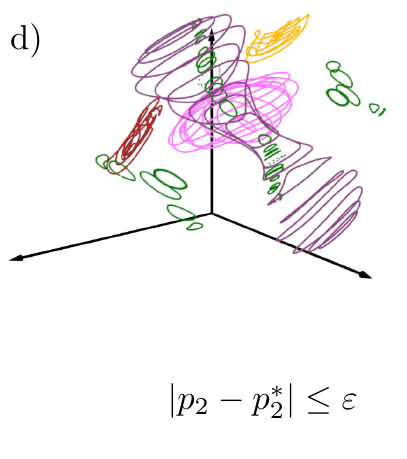}
  \caption{\label{fig:psslice-csm-varying-p2}
    \Psslcs{} for $|\p_2 - \p_2^*| \le \sectioneps$
    of two coupled standard maps
    at strong coupling $\xic = 1.0$ for
    varying $\p_2^* = 0.0, -0.0875, -0.14$, and $-0.1925$,
    \subfiga{} -- \subfigd.
    The color of the orbits is the same as in
    Fig.~\ref{fig:psslice-csm-strong-coupling}. %
    \MOVIEREFALT
  }
\end{figure*}

A global view of the dynamics in the full phase space is obtained by
varying the value of $\p_2^*$ of the slice
condition~\eqref{eq:slice-condition-in-coordinate}.
This is illustrated in the sequence of plots in
Fig.~\ref{fig:psslice-csm-varying-p2}, where
$\p_2^* = 0.0, -0.0875, -0.14, -0.1925$ is chosen
and only regular tori are shown.
Thereby, the slice condition $\p_2^*$ is shifted away from the fixed
point $(\p_2 = 0)$ and we observe that the visible regular region shrinks and
finally vanishes (not shown).
The outermost tori at $\p_2^* = 0$ (\colorregularbeyonda{} and
\colorregularbeyondb{}) persist the longest.
Both features can be understood from a \twoD{} analogy where a \oneD{}
slice of a regular island is shifted away from the fixed
point.
Furthermore, we observe the appearance of two other members of the
period-$7$ sub-region (\colorresonance{} and \colorresonanceb) in
Figs.~\ref{fig:psslice-csm-varying-p2}~\subfigb{} and \subfigd. Varying $\p_2^*$ to
positive values yields qualitatively the same sequence of pictures.

\label{text:3d-analogy-b}
While typically tori appear twice in the \psslc{}, under variation of $\p_2^*$
every pair of \oneD{} lines finally coalesces into a single \oneD{} line
before it disappears, see e.g.\ central \colorregularbeyonda{} tori in
Fig.~\ref{fig:psslice-csm-varying-p2}~\subfigd. This can again be understood
from an analogy in \threeD{} space when a \twoD{} section is shifted out
of a \twoD{} torus.

\subsection{\label{sec:slic-diff-degr}Slice condition with different
  coordinates}

Instead of choosing $p_2$ as the coordinate for the slice
condition~\eqref{eq:slice-condition-in-coordinate},
one can choose any of the other coordinates
to obtain a complementary view of the underlying phase space.
If, however, the system under consideration is strongly coupled, then
the single degrees of freedom $(\p_i, \q_i)$ are heavily
intertwined. Therefore, usually none of the slices along any of the $\q_i$ or
$\p_i$ has an advantage over any other possible choice.

Note that more generally one can define a rotated \psslc{}
by the slice condition
\begin{align}
  \label{eq:general-hyperplane}
  |\pointPS\cdot\vecsectionnormal - \sectiondist | &\leq \sectioneps,
\end{align}
for points $\pointPS$ in phase space, where $\vecsectionnormal$ is the
normal vector to the chosen slice and \sectiondist{} is the distance
of the slice to the origin.

\subsection{\label{sec:adapted-phase-space}Adapted 3D phase-space slices}

When focusing on
the vicinity of fixed points \fixedpoint{} or periodic orbits \periodicpoint{}
one can choose a more appropriate slice than the one given by
Eq.~\eqref{eq:slice-condition-in-coordinate}.
For this we use the properties of the linearized system at
\fixedpoint{} or \periodicpoint{} to define the phase-space slice.
For simplicity we will describe this approach for fixed points
as it trivially extends to periodic orbits.

\begin{figure*}
  \includegraphics{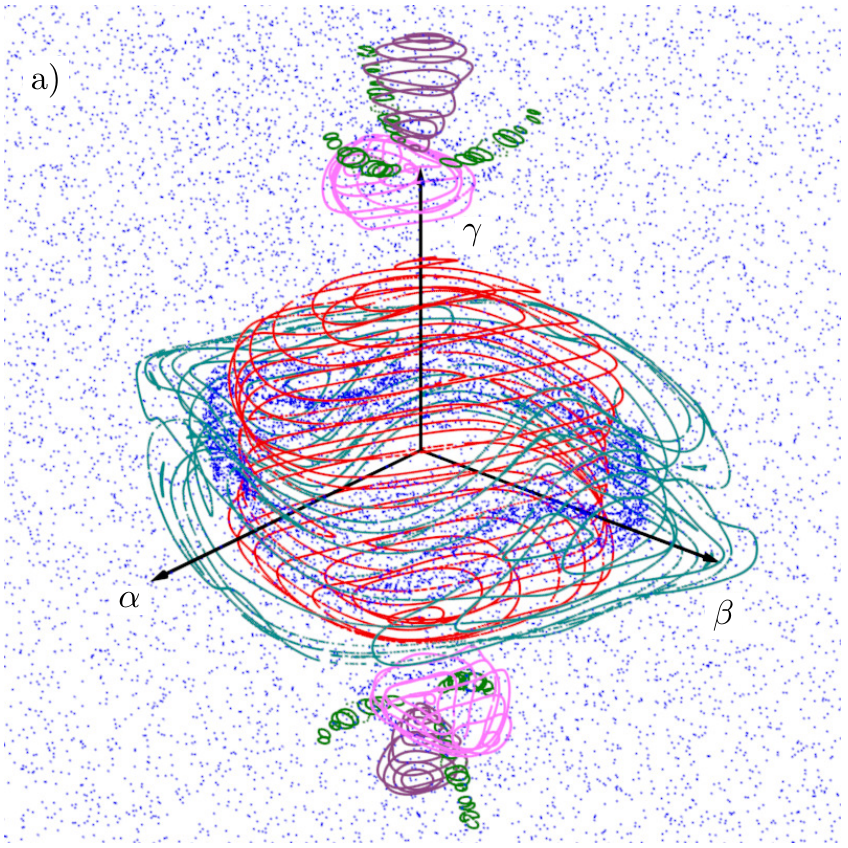}%
  \includegraphics{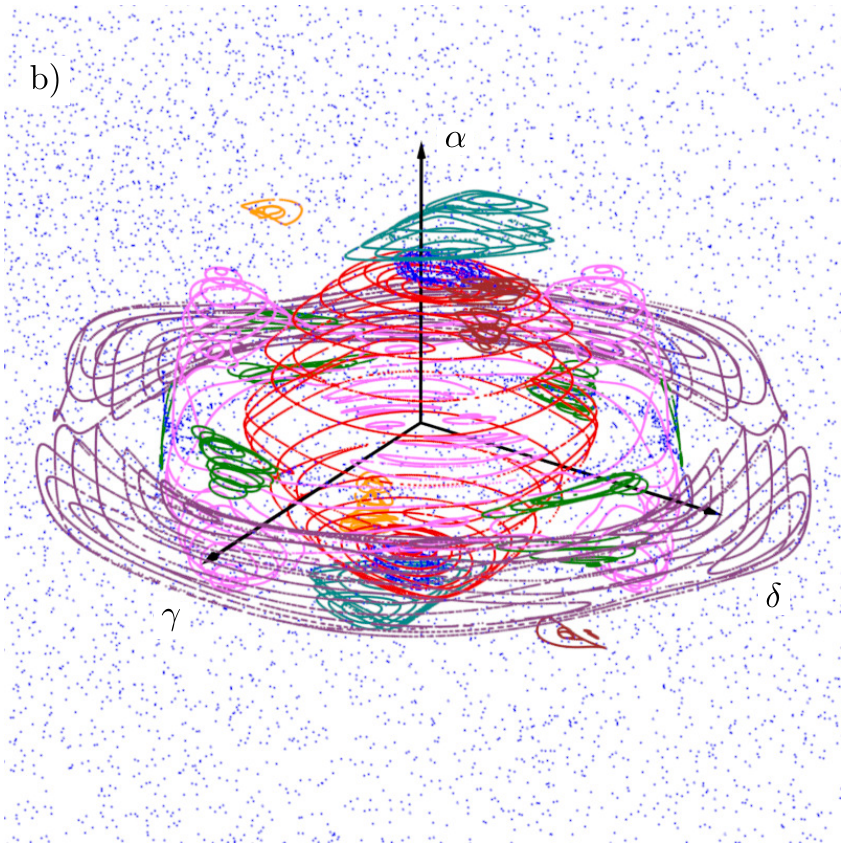}%
  \caption{\label{fig:skew-ps-section-for-csm} Adapted \psslcs{} of
    two coupled standard maps for the vicinity of the central
    fixed point \fixedpoint{}, for \subfiga{} coordinates $\alpha,
    \beta, \gamma$ and slice condition $|\delta| \leq \sectioneps$ and
    \subfigb{} coordinates $\alpha, \gamma, \delta$ and slice
    condition $|\beta| \leq \sectioneps$. %
    Shown are the orbits from
    Figs.~\ref{fig:psslice-csm-strong-coupling} and
    \ref{fig:psslice-csm-varying-p2}
    using the same color code. %
    \MOVIEREF }
\end{figure*}

For an elliptic-elliptic fixed point \fixedpoint{} the
four eigenvectors appear in pairs %
$(\vec{\DPevec}_1, \vec{\DPevec}_1^*)$ and %
$(\vec{\DPevec}_2, \vec{\DPevec}_2^*)$ %
corresponding to the eigenvalue pairs $(\DPeigenvalue_1,
\DPeigenvalue_1^*)$ and $(\DPeigenvalue_2,
\DPeigenvalue_2^*)$~\cite{LicLie92}.
The spaces spanned by the pairs
\begin{align}
  \label{eq:invariant_eigenspace_linearized_map}
  \overset{\displaystyle \vecealpha \propto
    \Re{\vec{\DPevec}_1}}{\underset{\displaystyle \vecebeta \propto \Im{\vec{\DPevec}_1}}{ }} \qquad\text{and}\qquad \overset{\displaystyle \vecegamma \propto \Re{\vec{\DPevec}_2}}{\underset{\displaystyle \vecedelta \propto \Im{\vec{\DPevec}_2}}{ }}
\end{align}
are invariant under the linearized dynamics.
In order to introduce a more appropriate slice for \fixedpoint{} we first
note that every point $\pointPS$ in phase space can be uniquely decomposed into
\begin{align}
  \label{eq:decomposition-4d-non-orthogonal}
  \pointPS = \alpha \vecealpha + \beta \vecebeta +
  \gamma \vecegamma + \delta \vecedelta
\end{align}
using the dual basis to the vectors from
Eq.~(\ref{eq:invariant_eigenspace_linearized_map}): Given the matrix
built column-wise from the vectors %
$\vecealpha, \vecebeta, \vecegamma$, $\vecedelta$ %
the dual basis vectors $\dualvecealpha, \dualvecebeta, \dualvecegamma,
\dualvecedelta$ can be obtained numerically via
\begin{align}
  \label{eq:dual-basis}
  \begin{pmatrix}
    \dualvecealpha | \dualvecebeta |
    \dualvecegamma | \dualvecedelta
  \end{pmatrix} & {} =
  {\begin{pmatrix}
    \vecealpha | \vecebeta |
    \vecegamma | \vecedelta
  \end{pmatrix}^{-1}}^{T}.
\end{align}
This provides the components %
$\alpha = \dualvecealpha \cdot \pointPS,
\beta = \dualvecebeta \cdot \pointPS,
\gamma = \dualvecegamma \cdot \pointPS$,
$\delta = \dualvecedelta \cdot \pointPS$.
Although the vectors $\vecealpha$, $\vecebeta$, $\vecegamma$,
$\vecedelta$ are non-orthogonal we can use any subset of them to
define an adapted \psslc: For example using the slice condition
\begin{align}
   |\delta - \sectiondist| \leq \sectioneps
\end{align}
the coordinates $\alpha, \beta$, and $\gamma$
can be plotted into an orthogonal \threeD{} coordinate system.
This slice is orthogonal to the dual vector $\dualvecedelta$.
Although this \threeD{} plot is not angle-preserving it displays the
features of the phase space in the vicinity of the elliptic-elliptic
fixed point \fixedpoint{} very clearly.

Moreover, these adapted \psslcs{} could also be used
to examine the vicinities of elliptic-hyperbolic
and hyperbolic-hyperbolic points.
Note that adapted \psslcs{} are equivalent to slices through the phase
space of a system after a normal form transformation. We now briefly
discuss two situations for using adapted \psslcs{}:
\paragraph{\label{sec:appl-rotat-phase}Elliptic-elliptic fixed points}

Firstly, we consider the central
elliptic-elliptic fixed
point \fixedpoint{}. Figs.~\ref{fig:skew-ps-section-for-csm}~\subfiga{} and~\subfigb{} show
the results for the slice conditions $|\delta| \leq \sectioneps$
and $|\beta| \leq \sectioneps$.
In Fig.~\ref{fig:skew-ps-section-for-csm}~\subfiga{}
the degree of freedom spanned by $(\vecealpha, \vecebeta)$ is completely visible
and in Fig.~\ref{fig:skew-ps-section-for-csm}~\subfigb{}
the one spanned by $(\vecegamma, \vecedelta)$ is completely visible.
Note that the slice conditions $|\gamma| \leq \sectioneps$
or $|\alpha| \leq \sectioneps$ lead to qualitatively
similar pictures, respectively.

Both Figs.~\ref{fig:skew-ps-section-for-csm}~\subfiga{} and~\subfigb{}
give a very clear representation of the regular phase-space region
where the central part (\colorregularcenter) shows \twoD{} tori as two
separate \oneD{} lines each. They are nicely stacked vertically on top
of each other as would be the case for an uncoupled, purely integrable
system. %
Figs.~\ref{fig:skew-ps-section-for-csm}~\subfiga{} and \subfigb{}
together provide a complementary visualization of the \fourD{} phase
space: Structures which are concentrated on the $\alpha$-$\beta$ plane
in Fig.~\ref{fig:skew-ps-section-for-csm}~\subfiga{} are localized
along the $\alpha$-axis in the other slice in
Fig~\ref{fig:skew-ps-section-for-csm}.~\subfigb{} and analogously
objects on the $\gamma$-$\delta$ plane lie on the $\gamma$ axis in
Fig.~\ref{fig:skew-ps-section-for-csm}~\subfiga{}. %
For example, the \colorregularbeyondb{} part of the regular region
(and also the blue chaotic orbit) in
Fig.~\ref{fig:skew-ps-section-for-csm}~\subfiga{} is above and below
in $\alpha$-direction in
Fig.~\ref{fig:skew-ps-section-for-csm}~\subfigb. This is true the
other way around for the \colorregularbeyonda{} and
\colorregularbeyondabent{} continuation of the central regular region
and the \colorresonancethreetower{} structure.

The adapted \psslcs{} yield very organized phase-space pictures, even
though their advantage is present mainly locally, close to the fixed
point used for defining the slice.
Fig.~\ref{fig:skew-ps-section-for-csm} shows that away from the
central elliptic-elliptic fixed point \fixedpoint{} the phase-space structures become
more involved, which is similar to \twoD{} maps if one moves outside
of the center of the main island (compare with
Fig.~\ref{fig:StdMaps}).
Also, not all phase-space structures have to be present, e.g.~the
stable vicinity of the period-$7$ orbit is shown in \colorresonance{}
in Fig.~\ref{fig:skew-ps-section-for-csm}~\subfigb{} but is not
visible in~\subfiga.

\begin{figure}

  \mbox{
  \includegraphics[width=4.25cm]{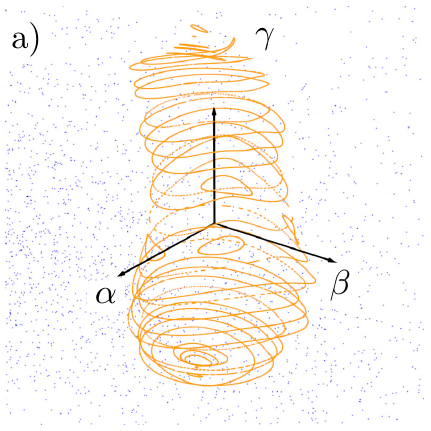}
  \includegraphics[width=4.25cm]{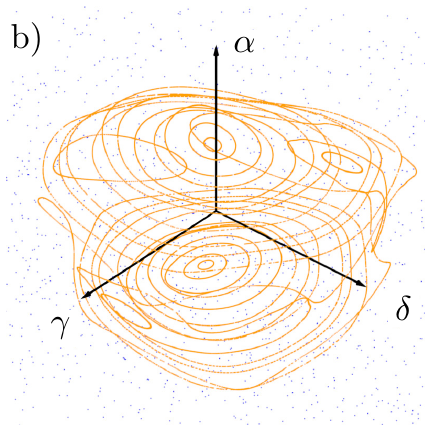}
  } 
  \caption{\label{fig:skew-ps-section-for-csm-period-7} Adapted
    \psslcs{} of two coupled standard maps for the vicinity of an
    elliptic-elliptic period-$7$ orbit, with slice condition
    \subfiga{} $|\delta| \leq \sectioneps$ and \subfigb{} $|\beta|
    \leq \sectioneps$. \MOVIEREF }
\end{figure}

\paragraph{\label{sec:appl-rotat-phase2}Elliptic-elliptic periodic orbits}

Suitably adapted phase-space slices can also be used to study periodic
orbits of higher period, in order to see the self-similar structure of
the phase space.
In Fig.~\ref{fig:psslice-csm-strong-coupling} the vicinity of an
elliptic-elliptic periodic orbit of period $7$ is shown in
\colorresonance. One of its points is located at $\periodicpoint =
(0.0, 0.0, 0.083438087, 0.118666288)$ and is therefore already visible
in Fig.~\ref{fig:psslice-csm-strong-coupling}. For this point the
adapted \psslc{} is shown in
Figs.~\ref{fig:skew-ps-section-for-csm-period-7}~\subfiga{}
and~\subfigb{}: The stable vicinity now looks very similar to the
central regular region in Fig.~\ref{fig:skew-ps-section-for-csm} as it
is again given by \oneD{} lines inside the \psslc{} vertically stacked
on top of each other. This is not obvious from the non-adapted
\psslcs{} shown in Fig.~\ref{fig:psslice-csm-strong-coupling}. One
could proceed from this $7$-fold map further down in the hierarchy of
periodic orbits.

Note that the fundamental frequencies of the tori shown in
Fig.~\ref{fig:skew-ps-section-for-csm-period-7} lie far away from the
frequency region shown in Fig.~\ref{fig:frequencies}. Taking the
frequencies with the largest amplitude leads to
$(\nu_1,\nu_2)=(\nicefrac{2}{7}, \nicefrac{1}{7})$ in
Fig.~\ref{fig:frequencies}.

\end{document}